# Local Convergence and Global Diversity: From Interpersonal to Social Influence[1]

Andreas Flache*, Michael W. Macy[+]

*Department of Sociology, University of Groningen, a.flache@rug.nl

[+]Department of Sociology, Cornell University, mwm14@cornell.edu

## *Abstract*

Axelrod (1997) prominently showed how tendencies toward local convergence in cultural influence can help to preserve cultural diversity if influence is combined with *homophily*, the principle that "likes attract." We argue that central implications of Axelrod's may change profoundly, if his model is integrated with the assumption of *social* influence as assumed by an earlier generation of modelers who did not use homophily, going back to French and Harary. Axelrod and all follow up studies employed instead the assumption that influence is *interpersonal* (dyadic). We show how the combination of social influence with homophily allows solving two important problems. As Axelrod noted himself, his model predicts cultural diversity in very small societies, but monoculture in large societies, which contradicts the empirical pattern. Our integration of social influence yields the opposite result, monoculture in small societies and diversity increasing in population size. The second problem was identified by Klemm et al. (2003a,b). They showed that cultural perturbation leaves only an extremely narrow window of noise levels in which diversity with local convergence can be obtained at all, and this window closes as the population size increases. Our model with social influence generates stable diversity with local convergence across a much broader interval of noise levels than models based on interpersonal influence.

## *1. Introduction*

Cultural diversity is both persistent and precarious. People in different regions of the world are increasingly exposed to global influences from mass media, internet communication,

1 This research has been supported by the Netherlands Organisation for Scientific Research, NWO (VIDI Grant 452-04-351), and by the U.S. National Science Foundation program in Human Social Dynamics (SES-0432917). Our work has benefited from stimulating discussions with Károly Takács, Michael Mäs, Tobias Stark, Tom Snijders, Christian Steglich and other members of the discussion group on norms and networks at the Department of Sociology of the University of Groningen.

interregional migration and mass tourism. English is rapidly becoming Earth's *Lingua Franca*, and Western music, films, and fashions can be found even in remote villages around the globe. While some authors point to the pervasiveness of cultural differences others expect globalization to lead to the emergence of a universal monoculture (cf. Greig, 2002). This expectation is grounded not only in empirical observation of recent trends but also in formal theoretical models of social influence that predict an inexorable tendency toward cultural convergence (French 1956; Harary 1959; Abelson 1964). Abelson and Harary proved that convergence to global consensus was inevitable in a connected (or what Abelson called a "compact") network – in which there are no components that are entirely cut off from outside influences. Obviously, diversity is guaranteed in a world where everyone is free to move to a small island that is isolated from outside influence, where they could live only with those identical to themselves. But Abelson wanted to know how diversity was possible in a worldwide culture, in which people cannot avoid exposure to those who are different. The problem, according to Abelson, stems from one of the fundamental principles of human interaction – social influence, or the tendency to alter one's opinions, attitudes, beliefs, customs, or other cultural traits to more closely resemble those of influential others (Festinger et al, 1950, cf. Axelrod 1997 for an overview). The models predict that cultural homogeneity is the inevitable long term outcome of processes by which individuals influence one another in response to the influences they receive. Later modifications in Friedkin and Johnsen's (1999) Social Influence Network Theory took into account that agents may still retain some residue of their original traits, no matter how great the influence of others. However, these models still imply that influence greatly reduces cultural differences, leading to a very high level of consensus under a broad range of conditions (cf. Friedkin 2001).

      This was the prevailing view among formal theories of cultural dissemination until about a decade ago. Building on previous work by Carley (1991), Axelrod (1997; see also Mark 1998) proposed an elegant extension of social influence models that incorporates *homophily*, or the "law of attraction" posited by Byrne (1969). This is the principle that "likes attract" (Lazarsfeld and Merton 1954; McPherson et al 2001). Like earlier formal models, Axelrod's assumed agents are connected in a spatial network. However, unlike earlier models, Axelrod assumed the strength of a tie between two neighbors could vary over time, depending on their similarity. Homophily generates a self-reinforcing dynamic in which similarity strengthens influence and influence leads to greater similarity. Axelrod's computational studies showed how this process can preserve global diversity. Once the members of two

cultural regions can no longer influence one another, their cultures[2] evolve along divergent paths. This model thus accounts for both tendencies that are evident in cultural evolution – on the one hand, the relentless swallowing up of cultural minorities, and at the same time, the inability for this process to end in monoculture.

Axelrod's breakthrough inspired a range of follow up studies, including Mark (1998, 2003); Shibanai et al. (2001); Greig (2002). Despite various modifications and extensions, the results supported Axelrod's basic conclusion – that cultural diversity can persist alongside local convergence[3].

Studies in the wake of Axelrod's seminal contribution have greatly advanced our understanding of the population dynamics of cultural dissemination. However, there is a potentially highly consequential hidden assumption in this line of research that goes back to Axelrod. The previous generation of models, dating back to French (1956), assumed that influence is a *social* phenomenon that can not be reduced to the interactions within a dyad that only comprises the source and the target of influence in a particular interaction. In this view, actors respond to the distribution of traits across all their neighbors, not just to the traits of the neighbor they actually interact with. Opinions are shaped in a field of simultaneous influences from multiple sources such that - all other things being equal – deviation from an opinion can be effectively precluded if the opinion is supported by a sufficiently large local majority of influential neighbors, as for example elaborated in Dynamic Social Impact Theory (Nowak, Szamrej & Latané, 1990). This reflects insights from empirical research on social influence. Latané and Wolf (1981) concluded that the social pressure on a target to adopt an opinion increases in the number of people that the target perceives to support the opinion, even if the supporters are not physically present in the interaction. In contrast, studies in the line of Axelrod assumed that influence was interpersonal rather than social[4]. Interpersonal influence occurs dyadically, between two people in a relationship, in isolation from others, even if these other network "neighbors" have much more in common with the agent. Social influence, on the other hand, is multilateral, involving all network neighbors simultaneously.

---

[2] Following Axelrod (1997), we use a very broad definition of "culture" that encompasses political opinions and beliefs, religious and moral values, artistic tastes in painting, music, fashion, cinema, etc. In short, "culture is taken to be what social influence influences" (Axelrod 1997, p. 207). He defines a cultural region as a set of contiguous cells with an identical cultural profile that is distinct from all neighbors.

[3] Two further important follow-up studies are Parisi et al (2003) and Centola et al (2006) both of which we will address in detail further below.

[4] The only expection we are aware of is Parisi et al (2003). However, their model differs from Axelrod's in a range of assumptions including no homophily, dichotomous traits, simultaneous updating on all features. They have not tested the separate effects of these deviations from Axelrod's model so that we can not learn from their study what the effects of social influence are.

We show in this paper that central implications of Axelrod's theory of cultural diversity change profoundly, when the assumption of social influence is integrated with the assumption of homophily within Axelrod's framework. We show in particular that including the assumption of social influence resolves two key problems with Axelrod's explanation of diversity that previous work has identified. The first problem is the inability of Axelrod's model to explain diversity in large populations. The second problem is the lack of robustness to noise.

Axelrod himself called attention to the critical limitation that his model can generate diversity only for very small populations. Axelrod's model predicts diversity in very small isolated societies composed of less than a few thousand people (at most) and monoculture everywhere else. This is a discouraging result because it is more plausible to expect monoculture in small isolated groups (such as intentional communities or remote tribal villages) and diversity in large societies, but Axelrod's model says it is the other way around, a discrepancy that concerned Axelrod as well (cf. p. 220).

Klemm et al. (2003a,b; for a recent overview see Centola et al. 2006) pointed to the problem of noise that ironically also seemed to lead to a solution to Axelrod's grid size problem. They relaxed Axelrod's assumption that cultural traits are entirely determined by influence from neighbors and allowed instead a small probability of random "perturbation" of cultural traits. They first showed that a small population that exhibits stable diversity under Axelrod's assumptions "drifts" toward monoculture in the presence of very small amounts of random cultural perturbation. Local convergence can trap a small population in an equilibrium in which influence is no longer possible because all neighbors are either identical or totally different. However, random cultural perturbations can disturb the equilibrium by generating cultural overlap between otherwise perfectly dissimilar neighbors, allowing social influence across cultural boundaries. This influence allows formerly dissimilar neighbors to become increasingly similar until no differences remain and a new cultural boundary forms around a larger region. Eventually this boundary too will be bridged by a perturbation that creates a common trait between otherwise dissimilar neighbors, and so on, until no differences remain.

Perturbations can also increase diversity by introducing cultural turbulence. If the rate of perturbation is sufficiently high, heterogeneity will be introduced faster than social influence can take advantage of the bridges created by perturbations to dissolve the boundaries between regions. As a consequence, a self-correcting equilibrium arises in which cultural regions keep changing their borders and their cultural identity, but the overall number and average size of culturally distinct regions fluctuates around a stochastically stable equilibrium level at which cultural perturbations introduces heterogeneity at a rate that balances the rate at which social influence makes neighbors more similar to each other.

Thus, perturbations have both a bridging effect which reduces diversity and a turbulence effect which increases it. Klemm et al. show that as population size increases, the turbulence effect swamps the bridging effect. This is an important result because it not only explains diversity in large societies but it also explains monoculture in small ones. The underlying mechanism is that if perturbations spread, it takes a much longer time in a large population before each member has been 'infected' than in a small population. As a consequence, higher population size increases the likelihood that at the time that some perturbation introduces new diversity, a previous perturbation still spreads elsewhere in the network. The result is sustained diversity in large populations, despite the 'pull' towards cultural homogeneity that follows from influence. This result led Klemm and his collaborators to suggest that in the "limit" of population sizes approaching infinity, "the fundamental idea of local convergence generating global polarization is recovered" (p. 67 045101-1, 2003a), but this time in the form of continuing cultural change.

Klemm et al's result shows that except for trivially small noise rates, the problem of cultural drift can only occur in small populations. For illustration, they reported that for a population size of 10.000 agents, cultural drift into monoculture only obtains when the probability that a cultural perturbation occurs within one social interaction is not larger than $10^{-5}$ (assuming F=10 features and Q=100 traits, cf p. 045101-3, 2003). Paradoxically, this implies that the mechanism of "network homophily" that Centola et al. (2006) have proposed to explain why cultural drift does not occur, is only needed to preserve diversity in small populations. But here – as Axelrod suggests – it may actually be less plausible. Centola et al add to Klemm's model the assumption that agents are free to choose their neighbors and will disconnect in particular from the influence of culturally dissimilar others. Then, homophily can generate a network with disconnected components in which every agent lives on a cultural island only with those exactly like themselves, as demonstrated by Centola et al. (2006).

But the main problem with the solution proposed by Klemm et al is not cultural drift. The problem is that under their assumptions cultural diversity with local convergence is highly fragile with respect to tiny changes in the amount of cultural perturbations. Cultural diversity with local convergence obtains only in a narrow window of perturbation rates, below which diversity collapses and above which local convergence is destabilized by evanescent fad-like behavior. Moreover, the size of the window closes down as population size increases. For large populations, the Klemm model predicts a world in which everyone marches to their own drum. A trivially small change in the noise rates shifts a large population from monoculture into the extreme opposite: cultural anomy with convergence neither at the local nor at the global level. For illustration, in the same condition for which Klemm et al reported monoculture below a probability of perturbation of $10^{-5}$, they also found that cultural regions dissolved into nearly isolated singletons when the rate of perturbation had increased to $10^{-4}$. Thus, monoculture is separated from cultural anomy by a difference of less than one divided by ten thousand in the likelihood that a perturbation occurs within one interaction event.

We show in this paper that an integration of social influence in Axelrod's and Klemm's models will solve both the grid size problem that Axelrod found in the absence of noise, and the lack of robustness of the explanation of cultural diversity offered by Klemm et al. The collapse of diversity in large populations that Axelrod found even in the absence of noise is driven by a lack of resistance to deviants. As Axelrod showed, the number of influence rounds needed until cultures have stabilized increases exponentially in the size of the population. But "the more time it takes for a territory to settle down, the more chance there is that ...regional boundaries will be dissolved." (p. XXX), because cultural changes may increase the similarity of previously disconnected neighbors on opposite sides of a regional boundary. However, we expect that social influence greatly increases the robustness of regional boundaries. Effective social influence across a boundary can only occur if a sufficient number of neighbors from the other side become similar to a focal agent. It takes much longer before this can happen than in the model with interpersonal influence. But in the meantime there is also ongoing pressure towards local conformity (in both models). As we will show, in the model with social influence this pressure is 'faster' than the pressure to reconnect to outside neighbors. As a consequence, the system settles down in a state of diversity even in a large population. The prevailing effect of population size is then simply that in a larger grid there are more local regions that form and settle down simultaneously than in a smaller grid.

Social influence similarly increases the robustness of cultural diversity in the presence noise. When agents are simultaneously influenced by multiple neighbors, rather than by just one randomly chosen neighbor at a time, it is no longer possible that a deviant lures its neighbors by influencing them one at a time. Moreover, isolated deviations from a local majority are unlikely to survive the conformity pressures from neighbors. With social influence, cultural perturbations may still create bridges between otherwise isolated regions, but the bridges will disappear sooner and have much less chance to be effective conduits of cultural influence across the regional border than under interpersonal influence. Cultural perturbations are also much less likely to lead to cultural anomy, because deviants are sooner brought back into the line of the locally dominant culture. The larger the neighborhood of agents, the stronger these stabilizing effects of social influence should be, because larger neighborhoods make isolated deviants an even smaller local minority. In sum, we argue that a model with social influence precludes both cultural drift into monoculture and cultural anomy, across a much larger interval of perturbation rates than Klemm et al found for their model, and that this difference in robustness increases when neighborhood size increases. To test this, we conducted computational experiments in which we compared a ceteris paribus replication of Axelrod's and Klemm's models with social influence to the original models with interpersonal influence, across a large range of noise levels.

      As an additional test of the robustness of social influence with homophily, we relaxed another hidden assumption introduced by Axelrod, that it is possible to have a zero probability of influence from maximally dissimilar neighbors and unit probability of influence from maximally similar neighbors. Human perception cannot be guaranteed to be perfect, thus there is always the possibility of 'selection error', we may see differences when none exist and see similarity when none exists. If we allow a positive probability of cultural perturbation (as Klemm et al. assume) as well as a positive probability of selection errors, this puts an additional pressure on cultural diversity. If a maximally dissimilar neighbor accidentally becomes influential, this creates a possibility that cultural influence occurs across the boundaries of two otherwise disconnected cultural regions, just as in the case of cultural perturbation. If a culturally identical neighbor fails to influence a target, this reduces the social pressure against outside influence and increases the likelihood that the focal agent adopts influences from cultural deviants. Thus, with selection error, cultural diversity should under the assumption of interpersonal influence be even less robust than with cultural perturbation alone. However, we expect that social influence also helps to solve this problem. With interpersonal influence, one selection error with a deviant neighbor is enough to 'infect' a conformist. With social influence, the deviant neighbor can not outvote the local majority.

In the section that follows, we describe extensions of the interpersonal influence models used by Axelrod and by Klemm et al. that introduce social influence as well as stochastic interaction. Section 3 presents results of computational experiments that show that social influence provides a more robust explanation for local convergence and global diversity, followed by a discussion of our work and future research perspectives in Section 4.

## *2. An Extended Model of Cultural Influence*

In the original Axelrod model, the population consists of *N* agents distributed over a regular bounded (non-toroidal) lattice, where each cell is occupied by a single agent who can interact only with the four neighbors to the N, S, E, and W (a von Neumann neighborhood). At any point in time, the cultural profile for an agent is a vector of *F* features. On any feature *f*, an agent has a trait *q* represented by an integer value in the range of $q = \{0...Q-1\}$, where *Q* is the number of possible traits on that feature. Formally, the cultural profile *C* of agent *i* is

$$C_i = (q_{i1}, q_{i2}, ..., q_{iF}), \quad q_{ix} \in \{0, 1, ..., Q-1\} \subset N_0. \tag{1}$$

In every discrete time step *t*, an agent *i* is randomly chosen from the population, and *i*'s cultural profile may then be updated through interaction with a randomly chosen neighbor *j*. The probability $p_{ij}$ that *i* and *j* will interact is given by the overlap in their cultural profiles, defined as the proportion of all *F* features with identical traits. Two agents *i* and *j* have identical traits on feature *f* if $q_{if} = q_{jf}$. If *i* and *j* interact, *j* influences *i* by causing *i* to adopt *j*'s trait on a feature randomly chosen from the set of features on which the two agents still differ from each other, i.e. $q_{if} \neq q_{jf}$.[5]

Following Klemm et al. (2003a,b), we then introduced cultural perturbation as a small probability *r* that a randomly chosen trait of an agent selected for interaction will be perturbed to a new value randomly chosen over the interval $\{0...Q-1\}$. More precisely, when an agent has been selected for a potential interaction with a neighbor, it is first decided whether interaction takes place and, in case it does, the interaction is conducted following Axelrod's rules. Subsequently, on a randomly selected feature of the focal agent, the current trait is perturbed with probability *r*. The possibility of a perturbation is also given if the focal agent did not interact. These are exactly the rules that Klemm et al (2003a,b) used.

---

[5] Note that Axelrod might have chosen f randomly from the set of all features, including those on which *i* and *j* have identical traits. Had he done so, the probability of interaction would increase linearly as the cultural distance between agents declines, but the probability of influence would decrease linearly. Axelrod's

We then relaxed two hidden assumptions in Axelrod's model – that influence is interpersonal (dyadic) and interaction is deterministic between agents who are perfectly similar or dissimilar.

I. *Social influence.* Instead of assuming that an agent can only be influenced by one neighbor at a time, we assume that an agent can be influenced simultaneously by all neighbors, depending on similarity. Having randomly chosen agent $i$ for possible updating, we then pick a neighbor $j$ who then becomes influential with a probability corresponding to $j$'s similarity with $i$, exactly as in Axelrod's model. However, instead of proceeding directly to the updating procedure, we then repeat, for all of $i$'s neighbors, this same procedure for selecting influential neighbors. Once the set of influential neighbors has been formed, the focal agent $i$ randomly selects a feature to update and then adopts the modal trait observed among the set of influential neighbors, with ties broken by random chance unless a tie occurs between the present trait of the agent and some other traits. In that case, the agent retains the present trait.

More precisely, if at a given timepoint $t$ an agent $i$ is randomly chosen for possible influence, then each of $i$'s neighbors $j$ is included by a random experiment into the subset of influential neighbors $I_i$ with a probability $p_{ij}$ that is given by the cultural overlap $o_{ij}$ between $i$ and $j$. Once the influence set has been identified, a feature $f$ is randomly selected from the set of features on which $i$ can be influenced that is: on which $i$'s trait differs from the trait of at least one member of the influence set. Agent $i$ observes the trait $q_{fk}$ on $f$ of every member $k$ of the set of influential neighbors of $i$. The modal trait, $m^*$, is the trait $m$ with the largest number of influential neighbors of $i$, $v_{fm}$, who have trait $m$ on feature $f$, or

$$\max_m v_{fm} = \|\{k \in I_i \mid q_{fk} = m\}\| \qquad (3)$$

If more than one trait satisfies this criterion and $i$'s current trait is amongst these traits, then $i$ retains the current trait. If $i$'s current trait is not amongst the traits satisfying the criterion, $i$ chooses randomly from among those traits.

This specification preserves the original interpersonal influence model as a condition nested within a more general form, such that social influence reduces to the dyadic model when an agent has only one influential neighbor. With more than one influential neighbor, it

---

specification insures that similarity affects only the probability of interaction and has no effect on the probability of influence, given that interaction occurs.

is no longer possible for a cultural innovator to induce a neighbor to adopt a perturbation that is an outlier within the distribution over all influential neighbors. The only way for an "unpopular" perturbation to spread is for the innovator to be the only neighbor included in the set of influential neighbors. Thus, social influence implies that perturbations tend to spread to a neighbor who has much in common with the innovator and little in common with other neighbors. In contrast, interpersonal influence implies that perturbations tend to spread to a neighbor who has much in common with the innovator, regardless of similarity with the other non-adopting neighbors.

II. Stochastic interaction. Following Axelrod, Klemm et al. assume that interaction always occurs between two agents with identical traits on all features and interaction never occurs between agents with dissimilar traits on all features. Klemm et al. introduce noise but they limit the effects to random errors in copying the trait of an influential neighbor, while preserving Axelrod's assumption that there are no errors in the decision to interact. We extended the effects of noise to include interaction errors as well as copying errors. Selection error is similar to random perturbation in that it can randomly alter the outcome of an interaction event. We assume that for every dyad, a selection error occurs independently of cultural perturbation with a probability $r'$. If by the normal selection procedure of the model a neighbor has been selected to be a member of the influence set of the focal agent, then a selection error enforces that this neighbor is removed from the influence set. If the neighbor has not been selected into the influence set, however, then the selection error will include the neighbor into the influence set. Both random events can potentially destabilize diversity. If a neighbor with zero overlap is included in the influence set due to selection error, this creates the possibility that cultural influence occurs across the boundaries of two otherwise disconnected cultural regions, just as in the case of cultural perturbation. If a culturally identical neighbor is excluded from the influence set, this reduces the social pressure against adopting outside influence and increases the likelihood that the focal agent adopts influences from cultural deviants. Technically, we implement perturbation and selection error as independent random events, but for both we assume the same error rate throughout all of the experiments reported in the remainder of this paper[6], i.e. $r=r'$.

---

[6] For comparison with the results of Klemm et al and Centola et al, we also replicated all experiments with only perturbation, i.e. r'=0. Nothing changes in our main result that diversity is vastly more robust with social influence than with interpersonal influence.

**3. Results**

We conducted two computational experiments to explore the effects of including social influence in Axelrod's model of cultural dissemination. With the first experiment, we wanted to know whether social influence can solve the grid size problem that Axelrod encountered without noise ($r = r' = 0$). With the second experiment, we tested the effect of different rates of noise ($r = r' > 0$) on cultural diversity both for interpersonal influence and for social influence. For a conservative test of the lack of robustness of diversity under interpersonal influence, we choose throughout our experiments a condition that Axelrod found to be conducive to high levels of diversity and that he used for most of the experiments reported in Axelrod (1997), $F=5$, $Q=15$. Moreover, we followed Axelrod and Klemm in assuming a rectangular cellular grid with boundaries (no torus) and letting every agent start out with a random cultural profile in which each feature is assigned a trait $q$ from a uniform distribution, such that each trait has an equal probability to be assigned. In both experiments, we manipulated as key condition interpersonal influence vs. social influence. In addition, we compared three populations with three different sizes, arranged on a rectangular grid structure. The smallest population we inspected consisted of $N=100$ agents arranged on a 10x10 cellular grid. We increased population size roughly by factor 10, inspecting in addition $N = 1024 = 32 \times 32$ and $N = 10.000 = 100 \times 100$. Per condition, we ran 10 independent replications and report average levels of diversity in the final iteration. Following Klemm et al, we report the size of the largest cultural region ($S_{max}$) as a measure of the (lack of) cultural diversity. A cultural region is defined as a set of contiguous cells with identical culture. The closer $S_{max}$ approaches the population size $N$, the less diversity we find in the population and the closer it is to one, the less local convergence occurs. For comparison across grid sizes, we report the share of the population that belongs to the largest region, i.e. $S_{max}/N$. Without noise, all runs finished when a static state was reached in which no more change was possible. This is the case when all cells have with each neighbor either zero overlap or are culturally identical to him. With noise, a static state is impossible. Moreover, it is hard to know when a dynamic equilibrium has been reached where the expected level of diversity is stable, particularly when the noise rate is low. Based on explorative tests, we fixed the duration of each simulation run close to $N*100.000$ iterations (10 million iterations for $N=100$, 100 million iterations for $N=1032$, and 1 billion iterations for $N=10.000$) and tested in some cases a tenfold number of iterations. This duration turned out to be long enough to reveal clear qualitative differences between interpersonal influence and social influence, but we have of course no certainty that levels of diversity remain qualitatively different when the number of iterations approaches infinity. However, the number of iterations we used is much higher than in some recent computational studies of stochastic process with a comparable number of agents (e.g. Baldassarri & Bearman 2007; Bruch and Mare, 2006), but corresponds to the standard in the study of models of cultural dissemination with noise (e.g. Centola et al, 2006).

*Experiment 1*

In experiment 1 we precluded noise ($r = r' = 0$) and compared the level of diversity in equilibrium for the three different population sizes. In the baseline condition, we assumed that smallest neighborhood size that Axelrod used, a von Neumann neighbor with 4 neighbors. This is a weak test of the fragility of diversity. The larger the neighborhood, the less likely it is that an agent can be entirely isolated from dissimilar others, hence the more likely that influence reduces cultural differences between regions. For interpersonal influence, Axelrod (1997) reported that larger neighborhoods greatly reduced cultural diversity. To assure that robust diversity may not just be an artifact of using too small neighborhoods, we also inspected neighborhoods with a radius of 6, where 'radius' refers to the largest number of steps in either vertical or horizontal direction that an agent can be separated from a neighbor. The largest number of neighbors with radius 1 is four neighbors, with radius 6 a neighborhood comprises up to 84 neighbors (except for the 10x10 grid, where only 74 neighbors are possible due to neighborhoods cutting across the borders of the grid).

Figure 1 reports the effects of population size and radius of interaction on normalized size of the largest region, broken down by dyadic influence and interpersonal influence. The figure confirms the discrepancy in the effects of population size on cultural diversity between the model with interpersonal influence and the model with social influence. For interpersonal influence, we find that diversity (normalized size of the largest region substantially smaller than 1.0) can only be sustained in the small population with N=100 and with a radius of 1. Both for N=1024 and N=10.000 we find virtual monoculture for both radiuses.

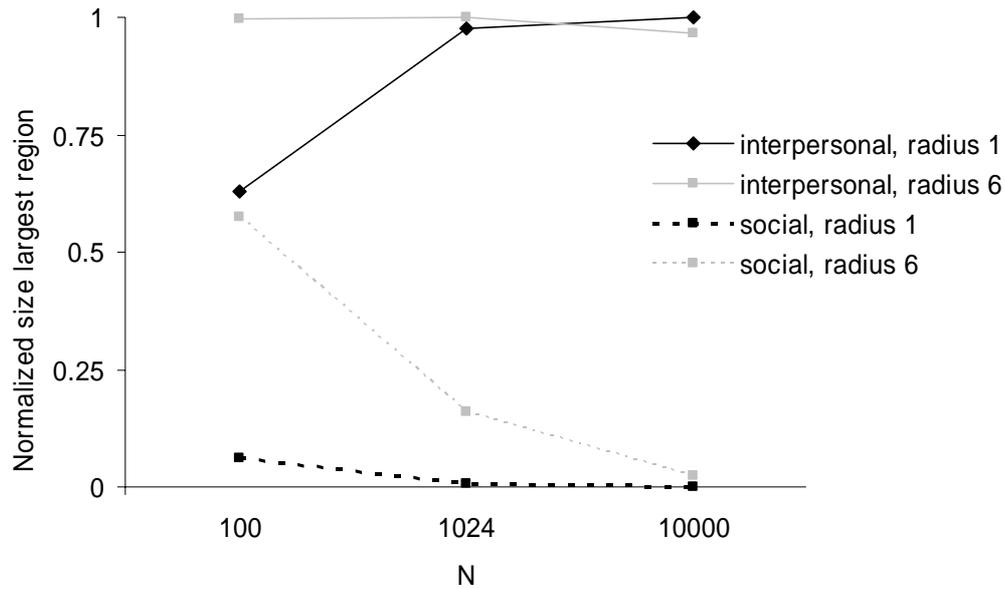

Figure 1. Normalized size of the largest region in equilibrium for three different population sizes (N) and two different interaction radiusus, broken down by interpersonal influence vs. social influence (F=5, Q=15, r=r'=0).

Figure 1 also clearly shows how social influence aligns the effect of population size with common sense intuition, contrary to Axelrod's original model. With the social influence model, we observed that the average size of the largest region declines when population size increases. For the larger radius of interaction of 6 steps, we found an average size of the largest cultural region of about 57% in a society of 100 agents, but this declined to about 16% for N=1024 and to about 2.2% for *N*=10.000. The qualitative effect of population size is robust across both interaction radiuses that we explored and the effect of radius is consistent across both models: larger radius increases the relative size of the largest cultural region. But none of the conditions we inspected yields an exact match of a pattern of monoculture in small societies and diversity in large societies. Even when the interaction radius is as large as 6 steps, we observe substantial diversity in the smallest society that we inspected (10x10). About 40% of the population are not member of the largest cultural region in this condition. As it turns out, adding noise aligns the social influence model closer with Axelrod's expectation.

*Experiment 2*

Klemm et al showed how small changes in the level of cultural perturbation can radically alter the level of cultural diversity generated by the interpersonal influence model, but they used different conditions in terms of the number of features (*F*) and traits (*Q*). Moreover, they tested only effects of cultural perturbation noise and their analysis was restricted to the smallest possible size of neighborhoods (radius 1). We focus in the following on neighborhoods with large size (radius 6) because this condition is the toughest test for the robustness of diversity under the social influence model[7]. As a first step, we wanted to know whether we can replicate for the interpersonal influence model Klemm's fragility of diversity to noise when we assume Axelrod's baseline condition (*F*=5,*Q*=15) and combine cultural perturbation and selection error in the model of noise ($r=r' > 0$). We varied the noise rate across a broad range from $r=r'=10^{-5}$ to $r=r'=10^{-1}$ and observed the effect on the relative size of the largest cultural region for the three different population sizes. Figure 2 reports the results that we obtained based on averages across the final states of 10 independent replications per condition. For clarity of the presentation, we scaled the noise level logarithmically.

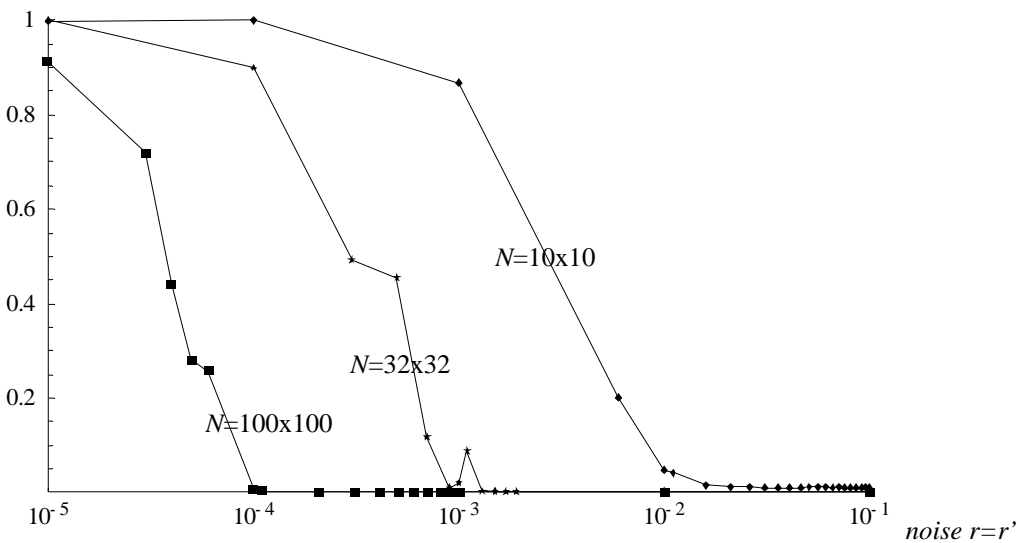

Figure 2. Interpersonal influence model. Effect of the level of noise on normalized size of the largest region ($S_{max}/N$) in the final iteration, for three different grid sizes (*F*=5, *Q*=15, radius=6).

---

[7] We also tested a radius of 1 and found no qualitative differences to the results obtained for radius 6.

Figure 2 shows that we can closely replicate the qualitative pattern that Klemm et al found for another set of parameters with the interpersonal influence model. For all three grid sizes, the initial diversity collapses into near monoculture ($S_{max}/N$ close to one) when the noise level is low, but only slightly higher noise levels suffice to radically change the outcome into near cultural anomy ($S_{max}/N$ close to zero). In between, there is a small window in which instable diversity goes together with local convergence. This window quickly narrows as grid size increases. For $N=10.000$, we found near monoculture at $r=r'=10^{-5}$ with a size of the largest region of 91.4% of the population, but when we increased the noise level by the tiny amount of 0.00004 ($r=r'=5*10^{-5}$), we already obtained a fair amount of instable diversity with a size of the largest region of about 44% and at $r=r'=2*10^{-4}$, near cultural anomy obtained with the largest region containing on average only 4.3 agents. For $N=1024$, we found monoculture for a somewhat broader range, between $r=r'=10^{-5}$ and $r=r'=10^{-4}$, but at $r=r'=3*10^{-4}$, monoculture had dissolved into diversity with local convergence, indicated by an average $S_{max}/N$ of about 50.5% and at $r=r'=1.5*10^{-3}$ the outcome was near cultural anomy with on average about 2.9 agents in the largest cultural region. Finally, for $N=100$, the range in which diversity with local convergence can be sustained is again somewhat broader than for the larger grids, but the width of the interval is still only about 0.01. Near monoculture occurred below $r=r'=10^{-3}$, but at $r=r'=10^{-2}$ the pattern had turned into near cultural anomy ($<S_{max}/N> \approx$ 0.05).

Next, we wanted to test whether social influence makes diversity more robust to noise. We conducted a ceteris-paribus replication of the experiment reported in figure 2, this time with the assumption that influence is social. Figure 3 shows a representation of the results.

Figure 3 demonstrates that social influence radically increases the stability of the level of cultural diversity against changes in the level of noise. For $N=10.000$ and $N=1024$, there is no qualitative change across the entire interval of noise levels that we inspected. At all levels of noise, we observe global diversity with local convergence. For $N=10.000$, the average size of the largest region ranges between $S_{max}=233.1$ and $S_{max}=608.3$ (about 0.023% and 0.061% of population size, respectively). The corresponding change for $N=1024$ is from $S_{max}=147.4$ to $S_{max}=341.4$ (about 14.4% to about 33.33%). For the small grid size, we find a qualitative change. Below a noise rate of $1.1*10^{-3}$, the outcome corresponds to diversity with local convergence. Above this level, the average outcome shifts to near monoculture with a size of the largest region of about 95% of the population size. In this region of the parameter space, we observe the precise pattern that Axelrod had expected, near monoculture in small societies and diversity in large societies. Across the entire noise range, the effect of population size is consistent with what we also obtained without noise (cf. figure 1). The larger the population, the higher is the level of diversity.

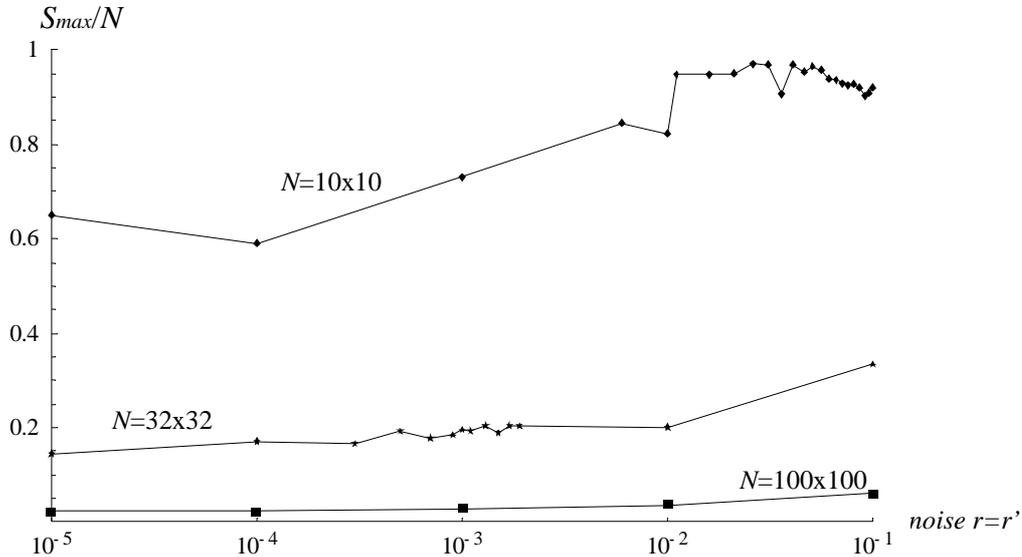

Figure 3. Social influence model. Effect of the level of noise on normalized size of the largest region ($S_{max}/N$) in the final iteration, for three different grid sizes ($F=5$, $Q=15$, radius=6).

Interestingly, the effect of noise on the level of diversity is with social influence not only much smaller than with interpersonal influence, but it is also in the opposite direction. The reason is that social influence is very effective in turning cultural innovators back into line. This prevents the fragmentation of cultural regions that higher rates of cultural perturbation entail under interpersonal influence. At the same time, higher noise also increases the probability of a selection error between agents from otherwise disconnected regions. As a consequence, the net effect of higher noise under social influence is that regions are more likely to merge and thus diversity declines. This also explains why we observe monoculture in small populations only when the level of noise is sufficiently high. Of course, when the noise rate is extremely high, perturbations eventually swamp the effect of influence. However, our results show that this effect does not occur for noise rates as high as $r=r'=0.1$.

If social influence implies that noise fosters pressures towards conformity rather than diversification, we may eventually obtain monoculture if we observe the dynamics of cultural dissemination for a sufficient number of iterations. To test this possibility, we increased the number of iterations tenfold to one million iterations per agent. We conducted a ceteris paribus replication of the experiment with the social influence model for the noise rate of $r=r'=10^{-3}$. We found that, as expected, for all three grid sizes the size of the largest region increased between 100.000 and 1.000.000 iterations per agent. However, the increase was only very small. For the 10x10 grid, the average of $S_{max}/N$ across 100 replications increased between iteration 10 million and 100 million from 82.24% to 93.6%. The corresponding figures for N=32x32 are an increase from 19% to 20% between 100 million and 1 billion iterations (based on 10 replications), and for $N$=100x100 we observed an increase from an average $S_{max}$ = 273.5 to 296 between 1 billion and 10 billion iterations (4 replications). These results suggest that while the process may indeed slowly move towards monoculture, we can consider diversity with local convergence to be temporarily stable for extremely long periods of time, certainly for larger populations.

## 4. Discussion and Conclusion

Axelrod (1997) showed how tendencies toward local convergence can actually help to preserve cultural diversity if interpersonal influence is combined with *homophily*, the principle that "likes attract." However, Axelrod's explanation suffers from two problems. As Axelrod noted himself, the model predicts cultural diversity in very small societies, but monoculture in large societies, which contradicts empirical observations. Klemm et al. (2003a,b) pointed to another problem that ironically turned out to lead to a solution of Axelrod's grid size problem. They showed that the stable cultural regions that Axelrod's model generates collapse into monoculture when very small amounts of "cultural perturbation" noise are introduced. But if a society is sufficiently large, the same small amount of cultural perturbation that generates monoculture in a smaller population, can yield a dynamic equilibrium with diversity and local convergence. Unfortunately, this result only holds for a very narrow range of perturbation rates, and this window closes as the population size increases. If the perturbation rate is too low, we fail to get global diversity and if it is too high we fail to get local convergence. We show that these problems can be resolved if Axelrod's and Klemm's approach is integrated with the assumption that influence is a truly social phenomenon, as assumed by an earlier generation of modelers going back to French and Harary, rather than assuming that influence is interpersonal, as Axelrod and Klemm did. We demonstrated that the combination of social influence with homophily generates diversity with local convergence that is much more robust to noise than in the models of Axelrod and Klemm, and that increases in group size both with and without noise.

Our work suggests that a simple but sociologically plausible extension of Axelrod's model may greatly improve the plausibility of model implications. But important questions remain. Our analysis depends on a range of simplifications that point to directions for future research. Earlier studies of social influence (French 1956; Abelson 1964) point to another assumption in Axelrod's model that is problematic. Axelrod assumes that all cultural features are nominal, like religion and language (hence represented as nominal scales). With nominal features, people are either identical or different, there are no shades of gray. However, it is immediately apparent that cultures can also be distinguished by metric features with *degrees* of similarity, such as the normative age of marriage or the enthusiasm for jazz. Even religion and language can form nested hierarchies in which some classes are closer to others (e.g., Congregationalists are closer to Unitarians than to Islamic Fundamentalists). In a previous study (Flache and Macy 2006b) we found that local convergence does not lead to global diversity in Axelrod's model if even a single cultural dimension is metric, no matter how many dimensions are nominal. Cultural homogeneity is then the ineluctable outcome, even when we assume away random mutation. Future work needs to explore whether the solution we propose in this paper also generalizes to the case of metric features, and to settings where diversity is threatened by a combination of metric features and noise.

Another simplification concerns the model of the elementary mechanisms. In particular, previous studies of cultural dissemination including the present paper have largely neglected the negative side of homophily and social influence – xenophobia and differentiation. Some exceptions are recent studies by Mark (2003) and Macy et al (2003), Flache and Macy (2006a) and Baldassarri and Bearman (2007) that allow for "negativity." Negativity may explain how agents remain different even when social interaction between them remains possible, but this has yet to be shown within the modeling framework of Axelrod's cultural dissemination model. We believe this is another promising avenue for future work.

Recently, Centola et al (2006) proposed network homophily as a solution to the problem of monoculture with small rates of cultural perturbation. We have argued that monoculture is only one part of the problem Axelrod and Klemm face, but we do not know whether network homophily can – like social influence - solve the problem that only slighter higher noise rates imply the extreme opposite of monoculture, cultural anomy. Finally, Parisi et al (2003) have proposed a model with social influence but without homophily that generates robust diversity even under cultural perturbation. Their model can not be compared to Axelrod's and Klemm's. We believe that future research needs to carefully explore if and under what conditions homophily is actually needed to explain diversity, once we also make the assumption that influence is social.